\documentclass[showpacs,prd,amsmath,amssymb]{revtex4}
\usepackage{graphicx}
\usepackage{epsfig}
\usepackage{graphics}
\usepackage{amsmath}
\usepackage{dcolumn}
\usepackage{bm}

\begin{document}

\title{Stability, causality and quasinormal modes of cosmic strings and cylinders}

\author{Alan B. Pavan}
\email{alan@fma.if.usp.br}
\author{E. Abdalla}
\email{eabdalla@usp.br}
\affiliation{Instituto de F\'{\i}sica, Universidade de S\~{a}o Paulo \\
C.P. 66318, 05315-970, S\~{a}o Paulo-SP, Brazil}

\author{C. Molina}
\email{cmolina@usp.br}
\affiliation{Escola de Artes, Ci\^{e}ncias e Humanidades, Universidade de
  S\~{a}o Paulo\\ Av. Arlindo Bettio 1000, CEP 03828-000, S\~{a}o
  Paulo-SP, Brazil}

%%%%%%%%%%%%%%%%%%%%%%%%%%%%%%%%%%%%%%%%%%%%%%%%%%%%%%%%%%%%%%%%%%%%%%%%%%%%%%%%%%%%%
\begin{abstract}
In this work we consider the evolution of a massive scalar field in cylindrically symmetric space-times. Quasinormal modes have been calculated for static and rotating cosmic cylinders. We found unstable modes in some cases. Rotating as well as static cosmic strings, i.e., without regular interior solutions, do not display quasinormal oscillation modes.  We conclude that rotating cosmic cylinder space-times that present closed time-like curves are unstable against scalar perturbations.
\end{abstract}

%04.30.Nk Wave propagation and interactions
%04.20.Ex Initial value problem, existence and uniqueness of solutions

\maketitle

%%%%%%%%%%%%%%%%%%%%%%%%%%%%%%%%%%%%%%%%%%%%%%%%%%%%%%%%%%%%%%%%%%%%%%%%
\section{Introduction}
%%%%%%%%%%%%%%%%%%%%%%%%%%%%%%%%%%%%%%%%%%%%%%%%%%%%%%%%%%%%%%%%%%%%%%%%

Cylindrically symmetric space-times, cosmic strings and cosmic cylinders in particular, are an important subclass of the full axial geometries, with a symmetry structure in many ways more complex than the spherically symmetric backgrounds. Although not compatible with asymptotic flatness, cylindrical solutions have been studied in many different physically meaningful contexts. Definitions and structural aspects have been investigated in \cite{carot,mikael}.
In cosmology and astrophysics they have been used to model the formation of structure \cite{Zeldovich}, gravitational lensing effect \cite{Gott} and gamma ray bursts \cite{gamma}. In condensed matter, they appear in the study of superconductivity and topological defects \cite{vilenkinbook}.

Beside formalism issues and phenomenological applications, cosmic strings display a variety of different phenomena.  They are important for the study of Einstein General Relativity in extreme situations, where the limits of the theory can be explored and  information about very new phenomena can be gathered. For example, they have been used to model time-machines \cite{deser,Jensen,Gott2}. Cylindrical space-times frequently have cuts in the angular variable. In case there is a rotation, one can imagine a voyager along the angular direction with
an effectively  superluminal velocity in a closed space-time path, hence a time machine. Thus, such geometries are generally associated with the possibility of traveling back in time, always a puzzling possibility \cite{Thorne,morris1,diaz1}. Hence the question whether these space-times are physically feasible. Effective superluminal velocities are not necessarily a pathology. For instance, in \cite{abdetal}, it has been shown that in braneworlds a gravitational sign can be sent through the bulk in a way that it can reach a distant point in the
brane faster than light propagating in the brane. However,
such signs seem to be irrelevant, because most of them advance, compared to a
light sign, by negligibly small amounts.

If we take causality as a fundamental principle, it is reasonable to conjecture that time machines cannot be constructed in the real world, although they are possible as exact solutions of Einstein equations. It is expected that these time machines do not survive as physical solutions, but we do not yet know any physical mechanism that prohibits the existence of such backgrounds.

In this work, we propose such mechanism, associating closed time-like curves to instabilities in certain cylindrical geometries. Taking the spherically symmetric cases as reference, not much has been stated about the stability of cosmic strings. This is specially true for rotating cosmic strings, which are relevant for time machine theoretical proposals. One exception, in the braneworld context, is the well known Gregory-Laflamme \cite{gregoryla} instability in the black strings.

We have calculated the evolution of a massive scalar field propagating in different cosmic string and cosmic cylinder backgrounds.  The relevant boundary conditions to be satisfied by a massive scalar field are presented. These conditions will be part of the quasinormal mode definition, which will be calculated considering these backgrounds. The static cases will be treated in the first place, because the rotating geometries present closed time-like curves (CTCs) and in these cases additional analysis is necessary to render the problem tractable. Analyzing the behavior of the field and their quasinormal modes we could present evidence of the stability of some of these space-times.

%%%%%%%%%%%%%%%%%%%%%%%%%%%%%%%%%%%%%%%%%%%%%%%%%%%%%%%%%%%%%%%%%%%%%%%%%%
\section{Cosmic string and cylinder space-times}
%%%%%%%%%%%%%%%%%%%%%%%%%%%%%%%%%%%%%%%%%%%%%%%%%%%%%%%%%%%%%%%%%%%%%%%%%%
The metric of a stationary and cylindrically symmetric space-time is given
by an expression of the form
\begin{eqnarray}
\label{mgeneral}
ds^2=-\tilde{F}\ dt^2+2\tilde{M}\ dtd\phi+\tilde{L}\ d\phi^2+\tilde{H}\ dr^2+\tilde{S}\ dz^2\quad,
\end{eqnarray}
where $(\tilde{F},\tilde{M},\tilde{L},\tilde{H},\tilde{S})$ are functions only of the radial coordinate $r$. All
space-times studied here are described by this Ansatz.

We start with the static cosmic cylinder background. It was obtained by Gott and Hiscock \cite{Gott, Hiscock} and its exterior solution is given by
\begin{eqnarray}
\label{scs1}
ds^2=-dt^2+dr^2+dz^2+\left(1-4\mu\right)r^2\ d\phi^2\quad,
\end{eqnarray}
where the coordinates range are $-\infty <t< \infty$, $\tilde{r}_{0}\sin(\theta_{M})\left(1-4\mu\right)^{-1}<r<\infty$, $0\leq \phi < 2\pi$, and $-\infty<z<\infty$. The interior solution is given by
\begin{eqnarray}
\label{c1}
ds^2=-dt^2+dr^2+dz^2+\bar{r}^2_{0}\sin(r/\bar{r}_{0})^{2}d\phi^2\quad,
\end{eqnarray}
with $-\infty < t < \infty$, $0 \leq r\leq \bar{r}_{0}\theta_{M}$, $0\leq \phi < 2\pi$, $-\infty<z<\infty$ and $\mu$ is the linear mass density. The matching point is defined by $r=r_{b}$ in the exterior radial coordinate and the corresponding interior radial coordinate is $r=r_{s}$. They are related by
\begin{eqnarray}
r_{b}=\frac{\bar{r}_{0}\sin(\theta_{M})}{1 - 4\mu}\quad, \qquad r_{s}=\bar{r}_{0}\theta_{M}\quad.
\end{eqnarray}
The other constants are related as
\begin{eqnarray}
\mu=\frac{1}{4}\left[1-\cos(\theta_{M})\right]\quad, \qquad \rho=\frac{1}{8\pi \bar{r}_{0}^2}\quad,
\end{eqnarray}
where $\rho$ is the energy density. The static cosmic string space-time can be obtained taking the limit $\bar{r}_{0}\rightarrow 0$.
For both static cosmic string and cosmic cylinder, we have assumed $\mu<1/4$. Additional details are presented in \cite{Gott}. The restriction on $\mu$ leads to
\begin{eqnarray}
\theta_{M}=\frac{r_{s}}{\bar{r}_{0}}<\frac{\pi}{2}\quad.
\end{eqnarray}
The rotating cosmic cylinder solution was obtained by Jensen and Soleng \cite{Jensen}. The exterior solution reads
\begin{eqnarray}
\label{cr1}
ds^2=-dt^2+dr^2+dz^2-8Jd\phi
dt+\left[(1-4\mu)^2(r+r_{0})^2-16J^2\right]d\phi^2\quad,
\end{eqnarray}
where $\mu$ is the linear mass density and $J$ the rate of angular
momentum per unit length. The constant $r_{0}$ determines the origin of
the exterior radial coordinate. For
\begin{eqnarray}
\label{cr2}
r+r_{0}<\frac{4|J|}{1-4\mu}
\end{eqnarray}
this space-time exhibits closed time-like curves.

Inside the cylinder there are at least two possible regular solutions. The first scenario considered in this work is the ``flower pot" model,
\begin{eqnarray}
\label{cr3}
ds^2=-\left[1+\frac{16J^{2}}{r^{4}_{s}}(r^{2}_{s}-r^{2})\right]dt^2-\frac{8Jr^{2}}{r^{2}_{s}}d\phi dt+r^{2}d\phi^2+dr^2+dz^2\quad,
\end{eqnarray}
 where $r_{s}$ is related to the exterior parameter  $r_{0}$ by
\begin{eqnarray}
\label{cr4}
r_{0}=\pm\sqrt{\frac{r^2_{s}+16J^2}{(1-4\mu)^2}}-r_{s}\quad.
\end{eqnarray}
Here there is no closed time-like curve.

The second type of internal solution treated is the so-called ``ballpoint pen" model. It describes gravity generated by the following distribution of matter,
\begin{eqnarray}
\label{cr5}
8\pi \rho=\lambda+\Omega^2\quad,  \qquad \Omega=\alpha\lambda(r_{s}-r)\quad,
\end{eqnarray}
where $\alpha\leq 1$ is an arbitrary constant which can be adjusted such that we may or may not have closed time-like curves. This interior solution is described by
\begin{equation}
\label{cr7}
ds^2=-dt^2-2Md\phi dt+\frac{\sin(\sqrt{\lambda}r)^2}{\lambda}-M^2d\phi^2+dr^2+dz^2\quad,
\end{equation}
\begin{equation}
M=2\alpha\left[(r-r_{s})\cos(\sqrt{\lambda}r)-
\frac{\sin(\sqrt{\lambda}r)}{\sqrt{\lambda}}+r_{s}\right]\quad.
\end{equation}
Imposing that exterior and interior solutions match continuously on the boundary $r_{s}$ of the rotating cosmic cylinder, the parameters are related by
\begin{eqnarray}
\label{cr8}
r_{0}=\left[\pm \frac{\sqrt{1-(1-4\mu)^2}}{(1-4\mu)
\arccos{(1-4\mu)}}-1\right]r_{s}\quad,\qquad \sqrt{\lambda}=\frac{\arccos\left(1-4\mu\right)}{r_{s}}\quad,
\end{eqnarray}
\begin{eqnarray}
\label{cr6}
\mu=\frac{1-\cos(\sqrt{\lambda }r_{s})}{4}\quad,\qquad J=\frac{2\alpha}{4}\left[ r_{s}-\frac{\sin(\sqrt{\lambda}r_{s})}{\sqrt{\lambda}}\right] \quad.
\end{eqnarray}

%%%%%%%%%%%%%%%%%%%%%%%%%%%%%%%%%%%%%%%%%%%%%%%%%%%%%%%%%%%%%%%%%%%%%%%%%%%%%%%%%%%%%%%%%%%%%%
\section{Evolution of a massive scalar field }
%%%%%%%%%%%%%%%%%%%%%%%%%%%%%%%%%%%%%%%%%%%%%%%%%%%%%%%%%%%%%%%%%%%%%%%%%%%%%%%%%%%%%%%%%%%%%%
The evolution of a minimally coupled massive scalar field $\Psi$ in a curved space-time is
given by the equation
\begin{eqnarray}
\label{pert1}
\Box \Psi(t,r,\theta,z)=\beta^2\Psi(t,r,\theta,z)\quad,
\end{eqnarray}
where $\beta$ is the mass of the field and the differential operator $\Box$ has the form
\begin{eqnarray}
\label{op}
\Box\Psi=\frac{1}{\sqrt{-g}}\partial_\mu\left(\sqrt{-g}
g^{\mu\nu}\partial_\nu\Psi\right)
\end{eqnarray}
and $g$ is the determinant of the metric. In the cylindrically symmetric space-time given by metric (\ref{mgeneral}) we have
\begin{eqnarray}
\label{eqgeneral}
-\tilde{L}\partial_{t}^2\Psi+\tilde{F}\partial_{\phi}^2\Psi+2\tilde{M}\partial_{t}
\left(\partial_{\phi}\Psi\right)+\frac{1}{2}\frac{\partial_{r}\left(\mathcal{X}
    \tilde{H}\tilde{S}\right)}{\tilde{H}^2\tilde{S}}\partial_{r}\Psi+\mathcal{X}\partial_{r}
\left(\frac{\partial_{r}\Psi}{\tilde{H}}\right)+\mathcal{X}\partial_{z}
\left(\frac{\partial_{z}\Psi}{\tilde{S}}\right)=\mathcal{X}\beta^2\Psi\quad ,
\end{eqnarray}
where $\mathcal{X}=\tilde{F}\tilde{L}+\tilde{M}^2$.

Since the space-time is stationary and invariant by translation in $z$ and rotations in $\phi$, the scalar field can be decomposed as
\begin{eqnarray}
\label{field}
\Psi(t,r,\phi,z) =  \sum_{m\in \mathbb{Z}} \int_{-\infty}^{\infty} \tilde{R}_{km}(t,r)\ e^{i(m \phi + kz)}\ dk\quad,
\end{eqnarray}
where $m$ is an integer number and $k$ is real. We will be interested in a quasinormal mode decomposition, and therefore we also consider a ``time-independent'' approach, based on a Laplace-like transform \cite{nollert}
\begin{eqnarray}
\label{field_qnm}
R(\omega,r) = \int_{0}^{\infty} \tilde{R}_{km}(t,r)\ e^{ i \omega t}\ dt\quad,
\end{eqnarray}
with $\omega$  extended to the complex plane. Substituting (\ref{field}) and (\ref{field_qnm}) in (\ref{eqgeneral}), we observe that this latter differential equation is separable and can be written as
\begin{eqnarray}
\label{eqpert}
\frac{1}{2}\frac{\partial_{r}\left(\mathcal{X} \tilde{H}\tilde{S}\right)}{\tilde{H}^2\tilde{S}}\partial_{r}R+
\mathcal{X}\partial_{r}\left(\frac{\partial_{r}R}{\tilde{H}}\right)=R\left[
\mathcal{X}(\beta^2+k^2)+\tilde{F}m^2-2\tilde{M}m\omega-\tilde{L}\omega^2\right]\quad.
\end{eqnarray}
The evolution of $\Psi$ in the exterior region of the cosmic cylinders, describing the vacuum, can be calculated straightforwardly. The radial evolution of the massive scalar field in the exterior of a rotating cosmic cylinder is given by
\begin{eqnarray}
\label{pertrotating}
x^2\frac{d^2R}{dx^2}+x\frac{dR}{dx}+R\left[ \left(
\omega^2-k^2-\beta^2\right)x^2-\frac{\left(m+4J\omega\right)^2}{(1-4\mu)^2}\right] = 0\quad,
\end{eqnarray}
where we defined the coordinate $x=(r+r_{0})$. Particular cases can be recovered: the static cosmic string and outside of the static cosmic cylinder setting $(J=r_{0}=0)$ and the rotating cosmic string setting $r_{0}=0$. If $\omega^2>k^2+\beta^2$ equation (\ref{pertrotating}) has an exact solution given in terms of Bessel functions. When $\omega^2<k^2+\beta^2$ equation (\ref{pertrotating}) has an exact solution given in terms of the modified Bessel functions decaying exponentially to zero at spatial infinity. Therefore, we will analyze only the condition $\omega^2>k^2+\beta^2$ since, to calculate the quasinormal modes, we want that the field behaves as an out-going traveling wave at spatial infinity.

For the interior of the static cosmic cylinder, the radial evolution of the scalar field has been obtained exactly. Applying (\ref{eqpert}) to the respective metric given by (\ref{c1}) we have
\begin{eqnarray}
\label{pertstatic}
\frac{d^2R}{dx^2}+\frac{1}{\tan(x)}\frac{dR}{dx}+R\left[ \left(
\omega^2-k^2-\beta^2\right)\bar{r}_{0}^2-\frac{m^2}{\sin^2(x)} \right]=0\quad,
\end{eqnarray}
where the radial coordinate $x = \bar{r}_{0}/r$ has been used. Equation (\ref{pertstatic}) has an exact solution in terms of the associated Legendre functions.

For the ``flower pot" model the function $\mathcal{X}$ and the radial derivative of the determinant of the metric $g$ are
\begin{eqnarray}
\label{pccri11}
\mathcal{X}=r^2\sigma_{1}^2=-g\quad,   \qquad
\sigma_{1}^2=\left(\frac{r_{s}^2+16J^2}{r_{s}^2}\right)\quad.
\end{eqnarray}
Substituting the functions above in (\ref{eqpert}) the resulting equation reads
\begin{eqnarray}
\label{pccri12}
r^2\frac{d^2R}{dr^2}+ r\frac{dR}{dr} +
R\left\{ \left[ \frac{1}{\sigma_{1}^2}\left(\omega-\frac{4Jm}
{r^2_{s}}\right)^2-k^2-\beta^2\right] r^2-m^2\right\} = 0\quad,
\end{eqnarray}
with Bessel functions as solutions.

For the ``ballpoint pen" model the function $\mathcal{X}$ and the determinant of the metric $g$ are
\begin{eqnarray}
\mathcal{X}=\frac{\sin^2(\sqrt{\lambda}r)}{\sqrt{\lambda}}=-g\quad.
\end{eqnarray}
Using (\ref{eqpert}) we obtain
\begin{eqnarray}
\label{pertsphero}
\sin^2(x)\frac{d^2R}{dx^2}+\sin(x)\cos(x)\frac{dR}{dx}
+R \left[ \frac{\left(\omega^2-k^2-\beta^2\right)\sin^2(x)}{\lambda}-
\left(m+2\alpha\omega f\right)^2\right]=0\quad,
\end{eqnarray}
with the function $f$ given by
\begin{eqnarray}
f= (r-r_{s})\cos(x)-\frac{\sin(x)}{\sqrt{\lambda}}+r_{s} \quad,
\end{eqnarray}
where $x=\sqrt{\lambda}r$. The equation (\ref{pertsphero}) can not be solved exactly. However, an approximate solution has been obtained in the quasi-static limit, defined as $\sqrt{\lambda}r_{s}\ll 1$. When such a limit is used in (\ref{cr6}), the angular momentum and the linear mass density become
\begin{equation}
J\sim 0+\mathcal{O}(x^3)\quad,\qquad \mu\sim\frac{\lambda r_{s}^2}{8}+\mathcal{O}(x^3)\quad,
\end{equation}
resulting in an exterior solution that behaves as a quasi-static solution. In this case equation (\ref{pertsphero}) turns into Bessel equation,
\begin{eqnarray}
\label{pertsphero2}
x^2\frac{d^2R}{dx^2}+x\frac{dR}{dx}+R\left[
\frac{\left(\omega^2-k^2-\beta^2\right)\ x^2}{\lambda}-
m^2\right] = 0\quad.
\end{eqnarray}
Another solution of (\ref{pertsphero}) can be obtained in the limit of large mass. The necessary condition to be satisfied in this limit depends on the interval of the radial coordinate. One can note that in the range $0 \leq r \leq r_{s}$ the function $f(r)$ is bounded. Therefore, the limit of large $m$ is reached when
\begin{eqnarray}
\label{largem}
m \gg \max_{r \in [0,r_{s}]} |2\alpha\omega\ f(r)| \quad,
\end{eqnarray}
and (\ref{pertsphero}) becomes
\begin{eqnarray}
\label{pertspherolargm}
\sin^2(x)\frac{d^2R}{dx^2}+\sin(x)\cos(x)\frac{dR}{dx}+R\left[
\frac{\left(\omega^2-k^2-\beta^2\right)\sin^2(x)}{\lambda}-m^2\right]
=0\quad.
\end{eqnarray}
This equation is solved by associated Legendre functions.

%%%%%%%%%%%%%%%%%%%%%%%%%%%%%%%%%%%%%%%%%%%%%%%%%%%%%%%%%%%%%%%%%%%%%%%%%%%%%%%%%%%%%%%%%%%%%%%%%%%%%%%%%%%%%%%%%%%%%%%%%%%%%%%%%%%%%%%%%%%%%%%%%%%%%%%
\section{Boundary conditions, Causality, Quasinormal modes}
%%%%%%%%%%%%%%%%%%%%%%%%%%%%%%%%%%%%%%%%%%%%%%%%%%%%%%%%%%%%%%%%%%%%%%%%%%%%%%%%%%%%%%%%%%%%%%%%%%%%%%%%%%%%%%%%%%%%%%%%%%%%%%%%%%%%%%%%%%%%%%%%%%%%%%%

\subsection{Boundary conditions}

In the static cosmic cylinder backgrounds, the space-time is divided in two different regions. Therefore, we need to specify junction and boundary asymptotic conditions. Namely, the scalar field in the interior region $\Psi^{in}$ must be regular at the origin, the field and its derivative must be continuous on the border $r_{s}$ of the cosmic cylinder and only outgoing waves can escape to infinity. Hence, these boundary conditions can be summarized as
\begin{eqnarray}
\label{bcstcs1}
|\Psi^{in}|&<&\infty \hspace{4,3cm} \textrm{when}\quad r\rightarrow 0\quad,\\
\nonumber\\
\label{bcstcs2}
\Psi^{in}&=&\Psi^{out},\qquad \frac{\partial\Psi^{in}}{\partial x^{\mu}}=\frac{\partial \Psi^{out}}{\partial x^{\mu}}\qquad \textrm{when}\quad r =  r_{s}\quad,\\
\nonumber\\
\label{bcstcs3}
\Psi^{out}&\sim& \frac{e^{-i\omega(t-r)}}{\sqrt{r}}\qquad\qquad \qquad\qquad \quad \textrm{when}\quad r\rightarrow \infty\quad,
\end{eqnarray}
where $\omega$ is a complex frequency of oscillation the scalar field. When the space-time  does not have a regular interior region, as the static cosmic string, only (\ref{bcstcs1}) and (\ref{bcstcs3}) are used as boundary conditions.
In the case of rotating space-times we need additional conditions to make the quasinormal mode problem well-posed. The difficulty is related to the fact that the space-times under consideration are not globally hyperbolic. Although this is not an insuperable obstacle, further considerations must be made \cite{Ishibashi, Thorne, wald, volovich}.

Specifically, there is a region $r<r_{ctc}$ where the temporal coordinate is not globally well defined because the vector fields $\partial/\partial t$ and $\partial/\partial \phi$ are time-like. A compact temporal coordinate $\phi$ leads to closed time-like curves.
To deal with this problem we follow the approach developed by Novikov, Thorne, Friedman and others, imposing a ``self-consistency principle''\cite{Thorne}, which roughly states that all events on CTCs influence each other around closed time-like lines in a self-adjusted way. One implementation of the self-consistency principle in the dynamical systems discussed in this article is done imposing that the propagation of scalar field  is periodic in the space-time regions that are identified. Labelling these regions by $\Gamma_i$ and $\Gamma_f$, we impose
$\Psi\left( \Gamma_i \right)=\Psi\left( \Gamma_f \right) $.
In the case of rotating cosmic cylinder space-times, this condition is satisfied in the causal region. What we need to do is to extent this condition to the non-causal region. We have
\begin{eqnarray}
\label{ctc}
\Psi(t,0,r,z)=\Psi(t,2\pi,r,z)\qquad \Longrightarrow \qquad \Phi(0)=\Phi(2\pi)\quad.
\end{eqnarray}
With this supplementary condition, the propagation of scalar fields become tractable.

%%%%%%%%%%%%%%%%%%%%%%%%%%%%%%%%%%%%%%%%%%%%%%%%%%%%%%%%%%%%%%%%%%%%%%%%%%%%%%%%%%%%%%
\subsection{Causality}
%%%%%%%%%%%%%%%%%%%%%%%%%%%%%%%%%%%%%%%%%%%%%%%%%%%%%%%%%%%%%%%%%%%%%%%%%%%%%%%%%%%%%%

In this section an analysis concerning causality is presented. The presence of closed time-like curves can be verified analyzing whether $g_{\phi\phi}<0$  somewhere. In our space-times, the non-causal region are infinitely long hollow cylinders.

The $g_{\phi\phi}$ components of the static cosmic string and cylinder are always positive-definite if $\mu<1/4$. Thus, the static cases can be declared causally well-behaved. The rotating cases need to be analyzed one by one.

\begin{figure}[!h]
\epsfig{file =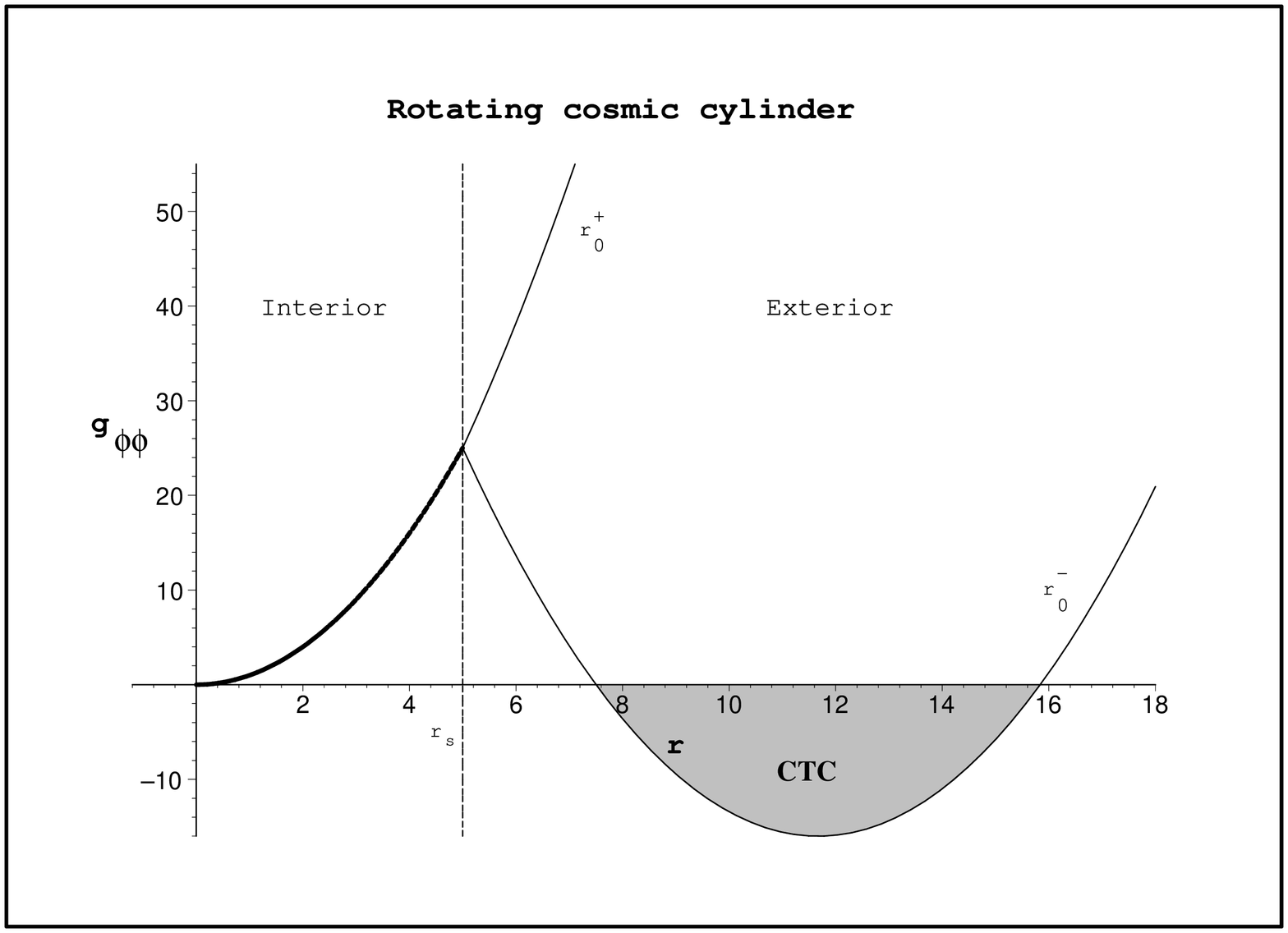, angle=-90, width=0.65 \linewidth}
\epsfig{file =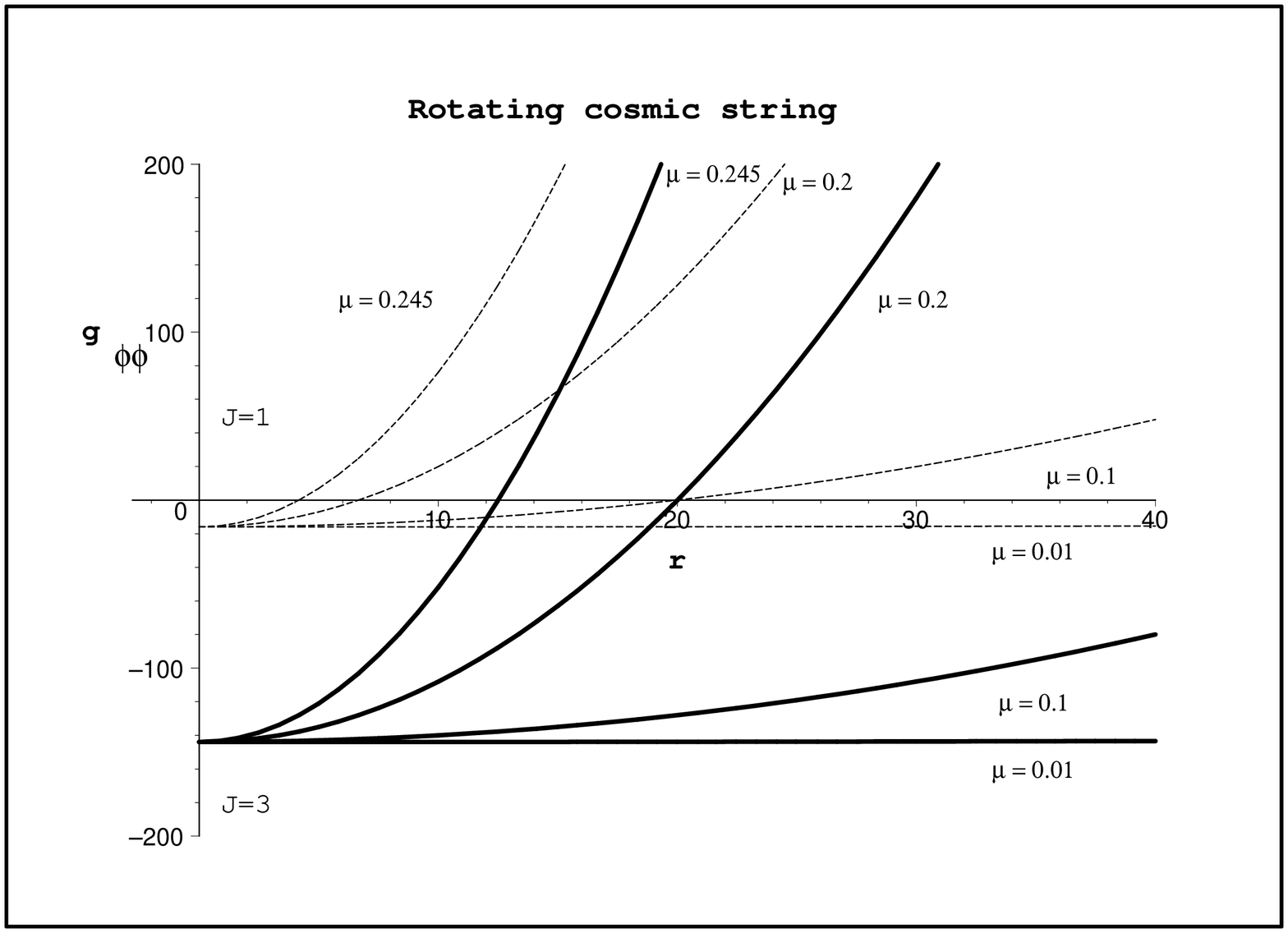, angle=-90, width=0.65 \linewidth}
\caption{The interior and exterior component $g_{\phi\phi}$ of the ``flower pot" model for $r_{0}^{+}$ and $r_{0}^{-}$ in (\ref{flcond}) with $r_{s}=5$ (top). The component $g_{\phi\phi}$ for the rotating cosmic string with different values of $J$ and $\mu$ (bottom).}
\label{figomega1}
\end{figure}

The rotating cosmic string admits a non-causal region with arbitrary size depending on the constants $\mu$ and $J$. In Fig. \ref{figomega1} one can see how these constants change the size of non-causal region. There is no restriction for values of $J$ but the linear mass density must be bounded.

\begin{figure}[!h]
\epsfig{file =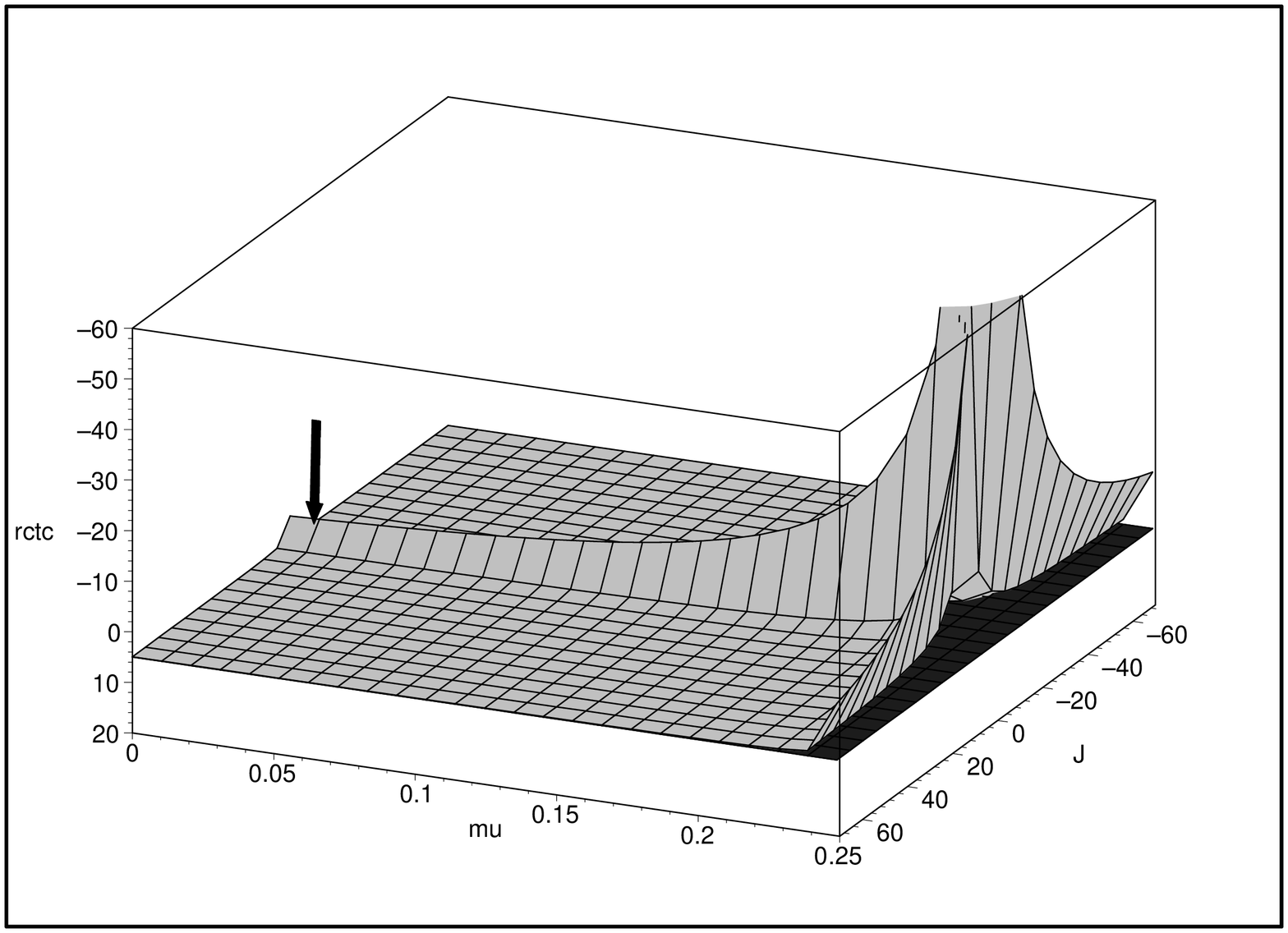, angle=-90, width=0.50 \linewidth}
\epsfig{file =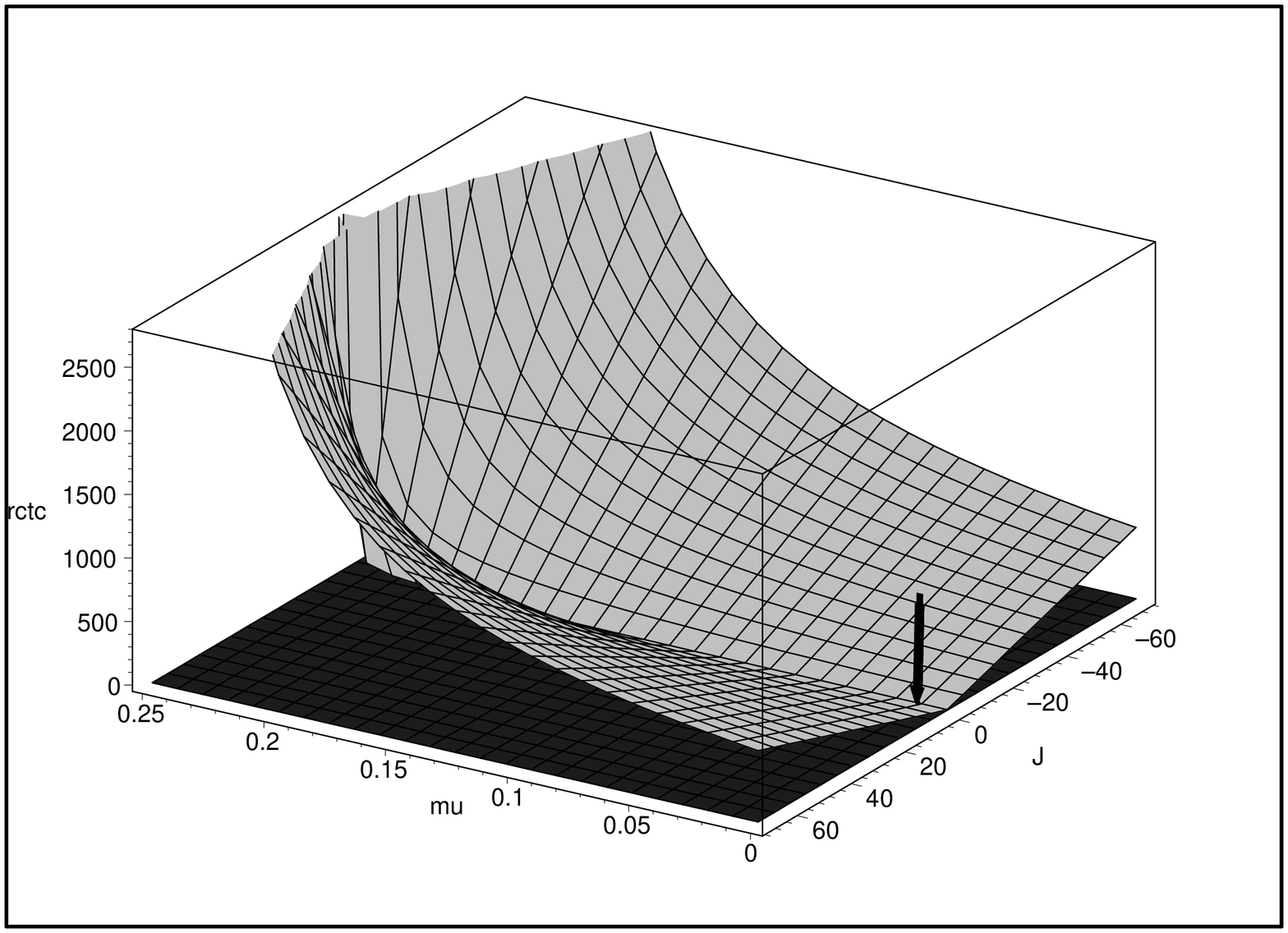,angle=-90, width=0.50 \linewidth}
\caption{Position of the non-causal limit surface $r_{ctc}$ in the rotating cosmic cylinder with ``flower pot" model for $r_{0}^{+}$ (top) and $r_{0}^{-}$ (bottom) in (\ref{flcond}). The radius of the cosmic string is $r_{s}=5$. The dark surface indicates the position of $r_{s}$ and the grey surface indicates the position of the non-causal limit surface $r_{ctc}$. The black arrow shows the position of the non-causal limit surface for $\mu=1.0 \times 10^{-2}$ and $J=1$.}
\label{paramflowerpot}
\end{figure}

The other two cases of rotating cosmic cylinders are more complex than the rotating cosmic string. For simplicity, we choose to preserve the interior region, which contains matter, against CTCs, reducing the range of the parameters. The exterior of the rotating cosmic cylinder may have CTCs.
In case of the ``flower pot" model the interior $g_{\phi\phi}$ component is always positive-definite, and therefore there is no CTC. In the exterior, the condition to have CTCs depends essentially on $r_{0}$ being given by
\begin{eqnarray}
\label{flcond}
r_{ctc}<\frac{\sqrt{16J^2}}{(1-4\mu)}-r_{0}\quad,\qquad \textrm{with}\ \ r_{0}=r_{0}^{\pm}=\pm\sqrt{\frac{r^2_{s}+16J^2}{(1-4\mu)^2}}-r_{s}\quad.
\end{eqnarray}
If $|r_{0}|$ is large enough $r_{ctc}<r_{s}$ resulting in the absence of a non-causal region. In Fig. \ref{paramflowerpot} we show the position of the non-causal limit surface $r_{ctc}$ for the rotating cosmic cylinder with the ``flower pot" model.

When the constant $r_{0}^{+}$ is chosen in (\ref{flcond}), the non-causal limit surface always lies before $r_{s}$, as seen in Fig. \ref{paramflowerpot} (top). In this case the exterior of the cosmic cylinder does not have CTCs. On the other hand, when the constant $r_{0}^{-}$ is chosen the non-causal limit surface always lies after of $r_{s}$ allowing CTCs. It is shown in Fig. \ref{paramflowerpot} (bottom). This conclusion is independent of the value of $r_{s}$. Therefore, in this case there is no restriction for the value of $J$. We still have $\mu<1/4$ with the additional constraint $\sqrt{\lambda}\ r_{s}<\pi/2$. Fig. \ref{figomega1} shows the causal behavior of the interior and exterior $g_{\phi\phi}$ components for the ``flower pot" model.

For the ``ballpoint pen" model, both the interior and exterior metrics allow the presence of CTCs. In this model a general analysis is quite difficult because of the number of free parameters. Each case must be analyzed independently, verifying whether the set of chosen constants satisfies given restrictions. The causal structure this space-time will depend on the adjustment of three constants $r_{s}$, $\lambda$ and $\alpha$. Analyzing the interior $g_{\phi\phi}$ component we can see that it can assume negative values. Recalling that we do not want CTCs in the interior solution, the set of constants must be chosen such that $g_{\phi\phi}$ is positive-definite. Again, the linear mass density must satisfy $\mu<1/4$ and $\sqrt{\lambda}r_{s}<\pi/2$. The angular momentum is fixed by the choice of $r_{s}$, $\lambda$ and $\alpha$.

\begin{figure}[!h]
\epsfig{file =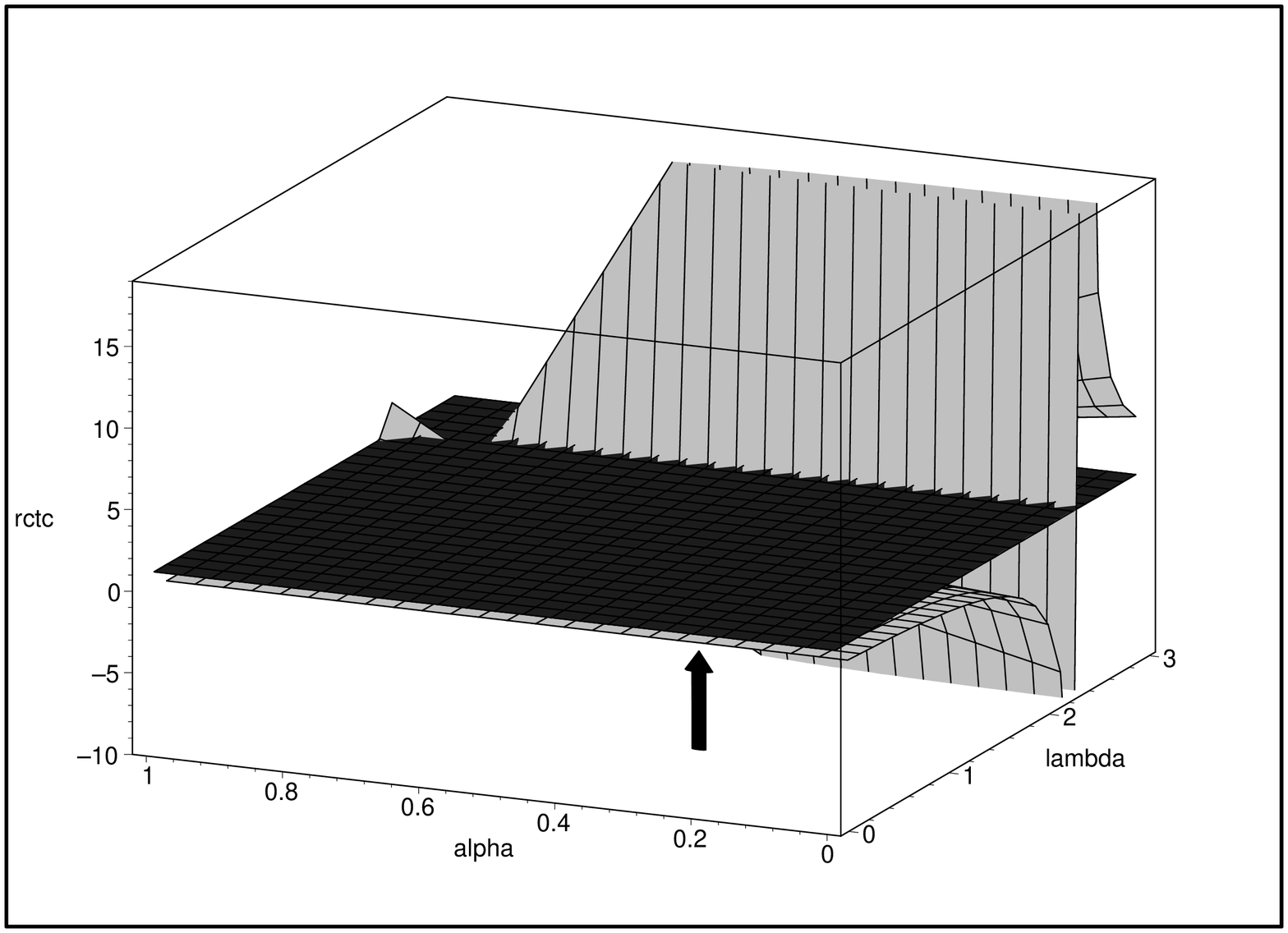, angle=-90, width=0.50 \linewidth}
\epsfig{file =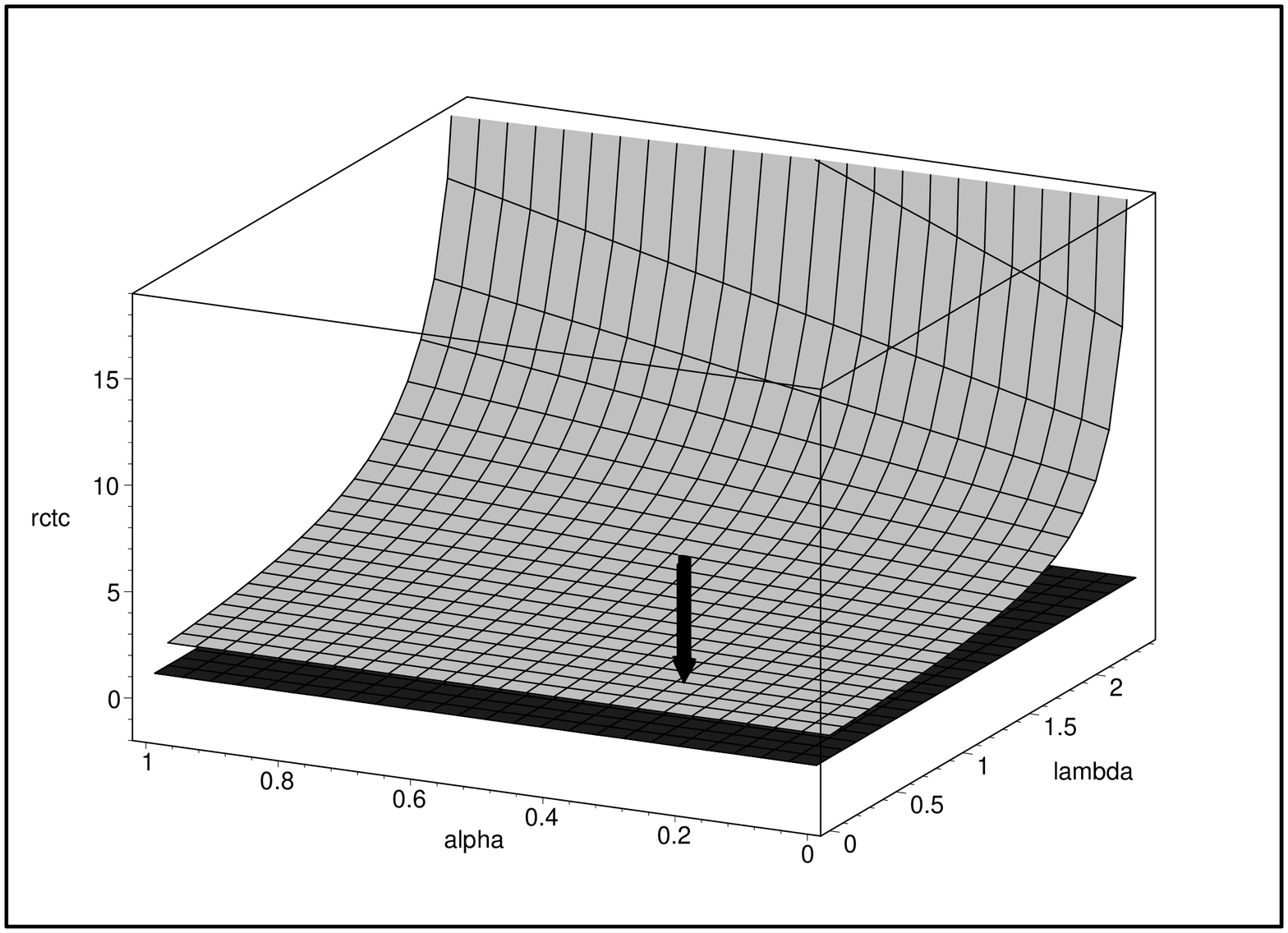,angle=-90, width=0.50 \linewidth}
\caption{Position of the non-causal limit surface $r_{ctc}$ in the rotating cosmic cylinder with ``ballpoint pen" model for $r_{0}^{+}$ (top) and $r_{0}^{-}$ (bottom) in (\ref{rctcballpoint}). The radius of the rotating cosmic cylinder is $r_{s}=1$. The dark surface indicates the position of $r_{s}$ and the grey surface indicates the position of the non-causal limit surface $r_{ctc}$. The black arrow shows the position of the non-causal limit surface for $\lambda=1.0\times 10^{-3}$ and $\alpha=0.2$.}
\label{rctcballpointstatic}
\end{figure}

As an example which captures the essence of the argument to be developed here, the set of constants $\lambda=1.0\times10^{-3}$, $\alpha=0.2$ and $r_{s}=1$ keeps the  interior $g_{\phi\phi}>0$, i.e. it is causally well-behaved. Now we need to analyze whether the set of chosen parameters allows the presence of CTCs in the exterior.
For the exterior of the ``ballpoint pen'' model, the condition to have CTCs also depends on $r_{0}$, being given by
\begin{eqnarray}
\label{rctcballpoint}
r_{ctc}<\frac{\sqrt{16J^2}}{(1-4\mu)}-r_{0}\quad,\qquad \textrm{with}\ \  r_{0}=r_{0}^{\pm}=\pm\frac{\sqrt{1-(1-4\mu)^2}\ r_{s}}{(1-4\mu)\arccos(1-4\mu)}-r_{s}\quad.
\end{eqnarray}
In Fig. \ref{rctcballpointstatic} the position of the non-causal limit surface $r_{ctc}$ for the rotating cosmic cylinder with ``ballpoint pen" model in quasi-static limit is shown.

\vspace{0.5cm}

When the constant $r_{0}^{+}$ is chosen in (\ref{rctcballpoint}), the position of the non-causal limit surface depends on the value of $\lambda$. For the particular case, $\lambda=10^{-3}$ and $\alpha=0.2$, $r_{ctc}$ is smaller than $r_{s}$ and in the exterior of this cosmic cylinder we do not find CTCs. If we choose $r_{0}=r_{0}^{-}$ in (\ref{rctcballpoint}) the position of the non-causal limit surface also depends of $\lambda$. But in this case $r_{ctc}$ is bigger than $r_{s}$ allowing CTCs. This considerations are general. Constraints on the values of $\lambda$ and $\alpha$ results in restrictions for $J$ and $\mu$. Therefore, the angular momentum and linear mass density can not assume arbitrary values in the ``ballpoint pen" model.

\newpage

\subsection{Quasinormal modes}

In this section we show the exact solutions for the evolution of the scalar field in the cosmic cylinder backgrounds. Quasinormal modes are numerically calculated.

\subsubsection{Static cosmic cylinder}

 The boundary conditions (\ref{bcstcs1}) and (\ref{bcstcs3}) result in
\begin{equation}
\label{bcstcs12}
R^{in}(r) = \tilde{C}_{1}\ P^{m}_{n}(\cos(r/\bar{r}_{0}))\quad,
\end{equation}
\begin{equation}
R^{out}(r) = \left\{
\begin{array}{lcl}
C_{1}\ H^{1}_{\nu}(pr) & \quad \textrm{when} \quad & \omega_{R}>0 \\
\\
C_{2}\ H^{2}_{\nu}(pr) & \quad \textrm{when} \quad & \omega_{R}<0
\end{array}
\right.
\quad ,
\end{equation}
where $n(n+1)=(\omega^2-k^2-\beta^2)\bar{r}_{0}^2$, $\nu^2=m^2/(1-4\mu)^2$ and $p^2=(\omega^2-k^2-\beta^2)$. The functions $R^{in}$ and $R^{out}$ are the radial part of the scalar field inside and outside of the cosmic cylinder. One can note that $R^{out}$ depends on the sign of the real part of $\omega$, due to the asymptotic behavior of Hankel functions. At the boundary, condition (\ref{bcstcs2}) implies
\begin{eqnarray}
\label{bcstcs11}
\frac{R'^{in}(r_{s})}{R^{in}(r_{s})}&-&\frac{R'^{out}(r_{b})}{R^{out}(r_{b})}=0\quad,
\end{eqnarray}
where $'$ denote a derivative with respect to $r$. The parameters are related by
\begin{eqnarray}
\label{bcstcs13}
r_{s}=\bar{r}_{0}\theta_{M}\quad,\qquad (1-4\mu)=\cos(\theta_{M})\quad,\qquad r_{b}=\frac{\bar{r}_{0}\sin(\theta_{M})}{(1-4\mu)}\quad,\qquad \rho=\frac{1}{8\pi \bar{r}_{0}^{2}}\quad.
\end{eqnarray}
If we choose $\rho$ and $r_{s}$, the other constants are fixed and we need only calculating the zeros of (\ref{bcstcs11}) for given $k,m,\beta$ to obtain the quasinormal frequencies $\omega$. Some quasinormal modes to static cosmic cylinder were numerically calculated. They are shown in Table \ref{tab1}.

\begin{table}[!h]
\centering
\caption{Quasinormal modes for static cosmic cylinder with the parameters $r_{s}=5.0$, $\rho=1.0\times 10^{-3}$, $\bar{r}_{0}=6.3$, $\mu=0.1$, $r_{b}=6.4$, $\beta=0$, $k=0$. (Left) Fundamental ($n=1$) and high overtone modes ($n>1$) for $m=0$. (Right) Fundamental mode ($n=1$) for several values of $m$.}
\label{tab1}
\begin{tabular}{cc}
\hline
$n$ & $\omega_{R}+i\omega_{I}$  \\
\hline
 1 & \ \  -0.66 - $i$ 0.54 \ \ , \ \  +0.66 - $i$ 0.54 \ \   \\

 2 & \ \  -1.33 - $i$ 0.65 \ \ , \ \  +1.33 - $i$ 0.65  \ \  \\

 3 & \ \  -1.98 - $i$ 0.72 \ \ , \ \  +1.98 - $i$ 0.72 \ \   \\

 4 & \ \  -2.62 - $i$ 0.78 \ \ , \ \  +2.62 - $i$ 0.78 \ \   \\

 5 & \ \  -3.25 - $i$ 0.82 \ \ , \ \ +3.25 - $i$ 0.82 \ \   \\

 6 & \ \  -3.89 - $i$ 0.85 \ \ , \ \ +3.89 - $i$ 0.85 \ \   \\

 7 & \ \  -4.52 - $i$ 0.88 \ \ , \ \ +4.52 - $i$ 0.88 \ \   \\

 8 & \ \  -5.15 - $i$ 0.91 \ \ , \ \  +5.15 - $i$ 0.91 \ \   \\
\hline
\end{tabular}
\hspace{0.5cm}
\begin{tabular}{cc}
\hline
$m$ & $\omega_{R}+i\omega_{I}$ \\
\hline
 0 & \ \ -0.66 - $i$ 0.54 \ \  , \ \ +0.66 - $i$ 0.54 \ \ \\

 1 & \ \ -0.21 - $i$ 0.42 \ \  , \ \ +0.21 - $i$ 0.42 \ \ \\

 2 & \ \ -0.49 - $i$ 0.51 \ \  , \ \ +0.49 - $i$ 0.51 \ \ \\

 3 & \ \ -0.76 - $i$ 0.58 \ \  , \ \ +0.76 - $i$ 0.58 \ \ \\

 4 & \ \ -1.01 - $i$ 0.63 \ \  , \ \ +1.01 - $i$ 0.63 \ \ \\

 5 & \ \ -1.26 - $i$ 0.68 \ \  , \ \ +1.26 - $i$ 0.68 \ \ \\

 6 & \ \ -1.51 - $i$ 0.72 \ \  , \ \ +1.51 - $i$ 0.72 \ \ \\

 7 & \ \ -1.76 - $i$ 0.76 \ \  , \ \ +1.76 - $i$ 0.76 \ \ \\
\hline
\end{tabular}
\end{table}

If we display the QNMs in the  $\omega_{I}\times\omega_{R}$ plane, we can see that the positive and negative $\omega_{R}$ of the quasinormal modes, are distributed symmetrically with respect to the $\omega_{I}$-axis. This happens for any value of $m$. The imaginary part, $\omega_{I}$, is always negative indicating stability of the static cosmic string against scalar perturbations. The $|\omega_{I}|$ oscillates in the four first modes and later grows, making  $\omega_{I}$ more negative. For $k\ne 0$ and $\beta\ne0$ the position of modes is pushed away of $\omega_{R}$-axis but the quasinormal spectrum qualitative behavior is the same.

\subsubsection{Rotating cosmic cylinder}

We now consider the rotating cosmic cylinder with the interior solution of the ``flower pot" model. Applying the boundary conditions (\ref{bcstcs1}), (\ref{bcstcs2}), (\ref{bcstcs3}) and (\ref{ctc}) to the scalar field we have
\begin{eqnarray}
\label{bcrtcs11}
\frac{R'^{in}(r_{s})}{R^{in}(r_{s})}-\frac{R'^{out}(r_{s})}{R^{out}(r_{s})}=0\quad.
\end{eqnarray}
The functions $R^{in}$ and $R^{out}$ are given by
\begin{equation}
\label{bcrtcs12}
R^{in}(r) = \tilde{C}_{1}\ J_{m}(p_{1}(r))\quad,
\end{equation}
\begin{equation}
R^{out}(r) = \left\{
\begin{array}{lcl}
C_{1} H^{1}_{\nu}(p(r+r_{0})) & \quad \textrm{when} \quad & \omega_{R}>0 \\
\\
C_{2} H^{2}_{\nu}(p(r+r_{0})) & \quad \textrm{when} \quad & \omega_{R}<0
\end{array}
\right.
\quad ,
\end{equation}
where $p_{1}^2=\frac{1}{\sigma_{1}^2}\left(\omega-4Jm/r_{s}^2\right)^2-k^2-\beta^2$, $m$ is an integer and $\nu^2=\left(m+4J\omega\right)^2/(1-4\mu)^2$. Once more the parameters associated to the space-time are related by
\begin{eqnarray}
\label{bcrtcs13}
r_{0}=r_{0}^{\pm}=\pm \sqrt{\frac{r_{s}^2+16J^2}{(1-4\mu)^2}}-r_{s}\quad,\qquad \mu <\frac{1}{4}\quad.
\end{eqnarray}
If we choose three constants, $\mu$, $J$ and $r_{s}$, the constant $r_{0}$ is fixed and then we need only to calculate the zeros of the equation (\ref{bcrtcs11}) for specific values of $k,m,\beta$ to obtain the quasinormal frequencies.

In Table \ref{tab2} some numerically computed frequencies are shown for rotating cosmic cylinder with ``flower pot" model. In contrast to the static case, here the distribution of $\omega_{R}$ is not symmetric with respect to $\omega_{I}$-axis. The imaginary part $\omega_{I}$ can assume negative or positive values depending on the values of $r_{0}^{-}$ in (\ref{bcrtcs13}) . The presence of modes with $\omega_{I}>0$ indicates instabilities of the scalar field perturbation.
On the other hand, if we choose the constant $r_{0}^{+}$ in (\ref{bcrtcs13}), these unstable modes are absent. In this case, the space-time can be considered stable. The relation between $r_{0}^{\pm}$ and the stability issues will be discussed in section \ref{final}.

\begin{table}[!h]
\centering
\caption{Quasinormal modes for rotating cosmic string with ``flower pot" model with the parameters $J=1.0$, $\mu=1.0\times10^{-2}$, $r_{s}=5.0$, $\beta=0$, $k=0$, $r_{0}=-12$ (top) and  $r_{0}=1.7$ (bottom). At the left, fundamental ($n=1$) and high overtone modes ($n>1$) for $m=0$. At right, fundamental mode ($n=1$) for several values of $m$. The $*$'s indicate the unstable modes.}
\begin{tabular}{ccc}
\hline
$n$ & \multicolumn{2}{c}{ $\omega_{R}+ i\ \omega_{I}$  }\\
\hline
1  & \ \ 0.064 + $i$\ 0.041 $^{*}$ \ \ & \ \ -0.064 + $i$\ 0.041 $^{*}$   \\

2  & \ \ 0.75 - $i$\ 0.23 \ \ \ \    & \ \ -0.21 + $i$\ 0.044 $^{*}$     \\

3  & \ \ 1.57 - $i$\ 0.32 \ \ \ \    & \ \ -0.79 + $i$\ 0.0082 $^{*}$     \\

4  & \ \ 2.39 - $i$\ 0.36 \ \ \ \    & \ \ -0.90 + $i$\ 0.014 $^{*}$     \\

5  & \ \ 3.20 - $i$\ 0.40 \ \ \ \    & \ \ -1.46 + $i$\ 0.031 $^{*}$     \\

6  & \ \ 4.01 - $i$\ 0.43 \ \ \ \    & \ \ -1.57 - $i$\ 0.32 \ \ \ \      \\
\hline
\hline
$n$ & \multicolumn{2}{c}{ $\omega_{R}+ i\ \omega_{I}$  }\\
\hline
1  & \ \ 0.46 - $i$\ 0.45 \ \ \ \ \ \ & \ \  -0.46 - $i$\ 0.45 \ \ \ \  \\

2  & \ \ 1.23 - $i$\ 0.57 \ \ \ \ \ \ & \ \  -1.23 - $i$\ 0.57 \ \ \ \ \\

3  & \ \ 2.02 - $i$\ 0.63 \ \ \ \ \ \ & \ \  -2.02 - $i$\ 0.63 \ \ \ \ \\

4  & \ \ 2.83 - $i$\ 0.67 \ \ \ \ \ \ & \ \  -2.82 - $i$\ 0.67 \ \ \ \ \\

5  & \ \ 3.63 - $i$\ 0.70 \ \ \ \ \ \ & \ \  -3.63 - $i$\ 0.70 \ \ \ \ \\

6  & \ \ 4.43 - $i$\ 0.73 \ \ \ \ \ \ & \ \  -4.43 - $i$\ 0.73 \ \ \ \ \\
\hline
\end{tabular}
\hspace{0.5cm}
\begin{tabular}{ccc}
\hline
$m$ & \multicolumn{2}{c}{ $\omega_{R}+ i\ \omega_{I}$  }\\
\hline
0  & \ \ 0.064 + $i$\ 0.041 $^{*}$      & \ \  -0.064 + $i$\ 0.041 $^{*}$   \\

1  & \ \ 0.026 - $i$\ 0.055 \ \ \ \     & \ \  -0.026 + $i$\ 0.034 $^{*}$   \\

2  & \ \ 0.051 + $i$\ 0.070 $^{*}$      & \ \  -0.56 + $i$\ 0.026 $^{*}$   \\

3  & \ \ 0.15 - $i$\ 0.22 \ \ \ \     & \ \  -0.19 + $i$\ 0.31 $^{*}$    \\

4  & \ \ 0.22 - $i$\ 0.31 \ \ \ \     & \ \  -0.37 + $i$\ 0.042 $^{*}$  \\

5  & \ \ 0.29 - $i$\ 0.39 \ \ \ \     & \ \  -0.44 + $i$\ 0.59 $^{*}$   \\
\hline
\hline
$m$ & \multicolumn{2}{c}{ $\omega_{R}+ i\ \omega_{I}$  }\\
\hline
0  & \ \  0.46 - $i$\ 0.45 \ \ \ \ & \ \  -0.46 - $i$\ 0.45 \ \ \ \ \\

1  & \ \  0.94 - $i$\ 0.39 \ \ \ \ & \ \  -0.83 - $i$\ 0.57 \ \ \ \ \\

2  & \ \  1.43 - $i$\ 0.39 \ \ \ \ & \ \  -0.12 - $i$\ 0.15 \ \ \ \ \\

3  & \ \  1.90 - $i$\ 0.40 \ \ \ \ & \ \  -0.07 - $i$\ 0.26 \ \ \ \ \\

4  & \ \  0.30 - $i$\ 1.90 \ \ \ \ & \ \  -0.17 - $i$\ 0.30 \ \ \ \ \\

5  & \ \  0.83 - $i$\ 2.09 \ \ \ \ & \ \  -0.26 - $i$\ 0.33 \ \ \ \ \\
\hline
\end{tabular}
\label{tab2}
\end{table}

For the rotating cosmic cylinder with the ``ballpoint pen" model, the quasinormal modes have been obtained in two different limits, namely quasi-static limit and large $m$ limit. Again the same procedure described before can be used to both limits, imposing the same boundary conditions.

The boundary conditions (\ref{bcstcs1}), (\ref{bcstcs2}), (\ref{bcstcs3}) and (\ref{ctc}) in the quasi-static limit give again the equation (\ref{bcrtcs11}). The functions $R^{in}$ and $R^{out}$ are given by
\begin{equation}
\label{bcrtcs12limit}
R^{in}(r) = \tilde{C}_{1}\ J_{m}(p_{1}(\sqrt{\lambda}r))\quad,
\end{equation}
\begin{equation}
R^{out}(r) = \left\{
\begin{array}{lcl}
C_{1}\ H^{1}_{\nu}(p(r+r_{0})) & \quad \textrm{when} \quad & \omega_{R}>0 \\
\\
C_{2}\ H^{2}_{\nu}(p(r+r_{0})) & \quad \textrm{when} \quad & \omega_{R}<0
\end{array}
\right.
\quad ,
\end{equation}
where now $p_{1}^2= \left(\omega^2-k^2-\beta^2\right)/\lambda$. The parameters associated to the space-time are related by
\begin{eqnarray}
\label{bcrtcs13limit}
r_{0}=r_{0}^{\pm}=\pm\frac{\sqrt{1-(1-4\mu)^2}\ r_{s}}{(1-4\mu)\arccos(1-4\mu)}-r_{s}\quad,\qquad \mu\sim\frac{\lambda r_{s}^2}{8} <\frac{1}{4}\quad, \qquad J\sim 0\quad.
\end{eqnarray}
Choosing $\lambda$ and $r_{s}$, the other constants are fixed and we need only to compute the zeros of (\ref{bcrtcs11}).

\newpage

\begin{table}[!h]
\centering
\caption{Quasinormal modes for rotating cosmic cylinder with ``ballpoint pen" model in quasi-static limit, with the parameters $J=2.0 \times 10^{-5}$, $\mu=1.0 \times 10^{-4}$, $r_{s}=1.0$, $\beta=0$, $k=0$, $\alpha= 0.2$, $\lambda=1.0 \times 10^{-3}$, $r_{0}= -2.0 $ (top) and $r_{0}=3.0 \times 10^{-4}$ (bottom). At left, fundamental ($n=1$) and high overtone modes ($n>1$) for $m=0$. At right, fundamental mode ($n=1$) for several values of $m$. The $*$'s indicate the unstable modes.}
\begin{tabular}{ccc}
\hline
 $n$ & \multicolumn{2}{c}{ $\omega_{R}+ i\ \omega_{I}$  }\\
\hline
 1  & \ \ 0.30 + $i$\ 0.41 $^{*}$ \   & \ \ \ \ -0.30 + $i$\ 0.41 $^{*}$ \ \ \\

 2  & \ \ 2.96 - $i$\ 0.94 \ \ \ \    & \ \ \ \ -2.96 - $i$\ 0.94 \ \ \ \ \ \ \\

 3  & \ \ 6.16 - $i$\ 1.28 \ \ \ \    & \ \ \ \ -3.83 + $i$\ 0.18 $^{*}$ \ \ \\

 4  & \ \ 9.33 - $i$\ 1.47 \ \ \ \    & \ \ \ \ -5.52 + $i$\ 0.17 $^{*}$ \ \ \\

 5  & \ \ 12.5 - $i$\ 1.62 \ \ \ \    & \ \ \ \ -6.16 - $i$\ 1.28 \ \ \ \ \ \ \\

 6  & \ \ 15.6 - $i$\ 1.73 \ \ \ \    & \ \ \ \ -7.01 + $i$\ 0.17 $^{*}$ \ \ \\
\hline
\hline
 $n$ & \multicolumn{2}{c}{ $\omega_{R}+ i\ \omega_{I}$  }\\
\hline
 1  & \ \ 2.61 - $i$\ 5.54 \ \ \ \ \ \    & \ \ \ \ -2.61 - $i$\ 5.54 \ \ \ \ \ \ \\

 2  & \ \ 5.94 - $i$\ 5.72 \ \ \ \ \ \    & \ \ \ \ -5.94 - $i$\ 5.72 \ \ \ \ \ \ \\

 3  & \ \ 9.17 - $i$\ 5.87 \ \ \ \ \ \    & \ \ \ \ -9.17 - $i$\ 5.87 \ \ \ \ \ \ \\

 4  & \ \ 12.4 - $i$\ 6.00 \ \ \ \ \ \    & \ \ \ \ -12.4 - $i$\ 6.00 \ \ \ \ \ \ \\

 5  & \ \ 15.5 - $i$\ 6.10 \ \ \ \ \ \    & \ \ \ \ -15.5 - $i$\ 6.10 \ \ \ \ \ \ \\

 6  & \ \ 18.7 - $i$\ 6.18 \ \ \ \ \ \    & \ \ \ \ -18.7 - $i$\ 6.18 \ \ \ \ \ \ \\
\hline
\end{tabular}
\hspace{0.5cm}
\begin{tabular}{ccc}
\hline
$m$ & \multicolumn{2}{c}{ $\omega_{R}+ i\ \omega_{I}$  }\\
\hline
0  & \ \ 0.30 + $i$\ 0.41 $^{*}$ \ \ \     & \ \ \ \   -0.30 + $i$\ 0.41 $^{*}$ \ \ \\

1  & \ \ 1.84 + $i$\ 0.25 $^{*}$ \ \ \     & \ \ \ \   -1.84 + $i$\ 0.25 $^{*}$ \ \ \\

2  & \ \ 0.34 + $i$\ 0.47 $^{*}$ \  \ \    & \ \ \ \   -3.06 + $i$\ 0.27 $^{*}$ \ \  \\

3  & \ \ 1.10 + $i$\ 0.70 $^{*}$ \ \ \     & \ \ \ \   -1.11 + $i$\ 0.70 $^{*}$ \ \  \\

4  & \ \ 1.92 + $i$\ 0.86 $^{*}$ \ \ \     & \ \ \ \   -1.92 + $i$\ 0.86 $^{*}$ \ \ \\

5  & \ \ 2.76 + $i$\ 0.99 $^{*}$ \ \ \     & \ \ \ \   -2.76 + $i$\ 0.98 $^{*}$ \ \  \\
\hline
\hline
$m$ & \multicolumn{2}{c}{ $\omega_{R}+ i\ \omega_{I}$  }\\
\hline
0  & \ \  2.61 - $i$\ 5.54 \ \ \ \ \ \      & \ \ \ \  -2.61 - $i$\ 5.54 \ \ \ \ \\

1  & \ \  1.07 - $i$\ 5.47 \ \ \ \ \ \      & \ \ \ \  -0.66 - $i$\ 5.46 \ \ \ \ \\

2  & \ \  2.74 - $i$\ 5.54 \ \ \ \ \ \      & \ \ \ \  -2.21 - $i$\ 5.60 \ \ \ \ \\

3  & \ \  1.00 - $i$\ 5.86 \ \ \ \ \ \      & \ \ \ \  -0.53 - $i$\ 5.90 \ \ \ \ \\

4  & \ \  2.37 - $i$\ 6.15 \ \ \ \ \ \      & \ \ \ \  -1.94 - $i$\ 6.27 \ \ \ \ \\

5  & \ \  0.85 - $i$\ 6.77 \ \ \ \ \ \      & \ \ \ \  -0.49 - $i$\ 6.81 \ \ \ \ \\
\hline
\end{tabular}
\label{tab3}
\end{table}

\begin{table}[!h]
\centering
\caption{Quasinormal modes for rotating cosmic cylinder with ``ballpoint pen" model in large $m$ limit with the parameters $J=0.1$, $\mu=1.0\times 10^{-2} $, $r_{s}=10$, $\beta=0$, $k=0$, $\alpha= 1.0$, $\lambda=1.0 \times 10^{-3}$, $r_{0}=-20 $ (top) and  $r_{0}=0.4 $ (bottom). At left, fundamental ($n=1$) and high overtone modes ($n>1$) for $m=70$. At right, fundamental mode ($n=1$) for several values of $m$. The $*$'s indicate the unstable modes and $\dagger$'s indicate the purely real modes.}
\begin{tabular}{ccc}
\hline
$n$ & \multicolumn{2}{c}{ $\omega_{R}+ i\ \omega_{I}$  }\\
\hline
1 & \ \  2.20  $^{\dagger}$         & \ \  -2.20  $^{\dagger}$ \ \   \\

2 & \ \  2.47 + $i$\ 3.99 $^{*}$ \ \ \    & \ \ \ \  -2.31 + $i$\ 4.30 $^{*}$ \ \  \\

3 & \ \  2.80 + $i$\ 3.86 $^{*}$ \ \ \   & \ \ \ \  -2.84 + $i$\ 4.08 $^{*}$ \ \  \\

4 & \ \  2.96 + $i$\ 3.78 $^{*}$ \ \ \   & \ \ \ \  -3.02 + $i$\ 3.99 $^{*}$ \ \   \\

5 & \ \  3.12 + $i$\ 3.70 $^{*}$ \ \ \   & \ \ \ \  -3.20 + $i$\ 3.90 $^{*}$ \ \   \\

6 & \ \  3.29 + $i$\ 3.61 $^{*}$ \ \ \   & \ \ \ \  -3.38 + $i$\ 3.80 $^{*}$ \ \   \\
\hline
\hline
$n$ & \multicolumn{2}{c}{ $\omega_{R}+ i\ \omega_{I}$  }\\
\hline
1 & \ \  0.033 - $i$\ 5.00 \ \ \ \ \ \  & \ \ \ \  -0.15 - $i$\ 4.70 \ \ \ \ \ \ \\

2 & \ \  0.21 - $i$\ 5.00 \ \ \ \ \ \  & \ \ \ \  -0.32 - $i$\ 4.70 \ \ \ \ \ \  \\

3 & \ \  0.39 - $i$\ 4.99 \ \ \ \ \ \  & \ \ \ \  -0.48 - $i$\ 4.69 \ \ \ \ \ \  \\

4 & \ \  0.57 - $i$\ 4.98 \ \ \ \ \ \  & \ \ \ \  -0.64 - $i$\ 4.67 \ \ \ \ \ \  \\

5 & \ \  0.75 - $i$\ 4.96 \ \ \ \ \ \  & \ \ \ \  -0.81 - $i$\ 4.66 \ \ \ \ \ \  \\

6 & \ \  0.93 - $i$\ 4.94 \ \ \ \ \ \  & \ \ \ \  -0.97 - $i$\ 4.63 \ \ \ \ \ \  \\
\hline
\end{tabular}
\hspace{0.5cm}
\begin{tabular}{ccc}
\hline
$m$ & \multicolumn{2}{c}{ $\omega_{R}+ i\ \omega_{I}$  }\\
\hline
60  & \ \ 1.88 $^{\dagger}$  \ \            &  \ \ -1.88 $^{\dagger}$ \ \ \\

62  & \ \ 1.94 $^{\dagger}$  \ \            &  \ \ -1.94 $^{\dagger}$ \ \ \\

65  & \ \ 2.04 $^{\dagger}$  \ \            &  \ \ -2.04 $^{\dagger}$ \ \ \\

67  & \ \ 2.10 $^{\dagger}$  \ \            &  \ \ -2.10 $^{\dagger}$ \ \ \\

70  & \ \ 2.20 $^{\dagger}$  \ \            &  \ \ -2.20 $^{\dagger}$ \ \ \\

72  & \ \ 2.26 $^{\dagger}$  \ \            &  \ \ -2.26 $^{\dagger}$ \ \ \\
\hline
\hline
$m$ & \multicolumn{2}{c}{ $\omega_{R}+ i\ \omega_{I}$  }\\
\hline
60  & \ \ 0.18 - $i$\ 4.30 \ \ \ \ \ \    &  \ \ \ \ -0.10 - $i$\ 4.05 \ \ \ \  \\

62  & \ \ 0.19 - $i$\ 4.44 \ \ \ \ \ \    &  \ \ \ \ -0.11 - $i$\ 4.18 \ \ \ \  \\

65  & \ \ 0.11 - $i$\ 4.65 \ \ \ \ \ \    &  \ \ \ \ -0.045 - $i$\ 4.38 \ \ \ \  \\

67  & \ \ 0.11 - $i$\ 4.79 \ \ \ \ \ \    &  \ \ \ \ -0.057 - $i$\ 4.51 \ \ \ \  \\

70  & \ \ 0.033 - $i$\ 5.00 \ \ \ \ \ \    &  \ \ \ \ -0.16 - $i$\ 4.71 \ \ \ \  \\

72  & \ \ 0.039 - $i$\ 5.14 \ \ \ \ \ \    &  \ \ \ \ -0.17 - $i$\ 4.84 \ \ \ \  \\
\hline
\end{tabular}
\label{tab4}
\end{table}

Quasinormal modes for the ``ballpoint pen" model in the quasi-static limit were obtained numerically being shown in the Table \ref{tab3}. The imaginary part $\omega_{I}$ can assume negative or positive values depending on the values of $r_{0}^{\pm}$ in (\ref{bcrtcs13limit}), as in the ``flower pot" model.
In the stable case, the first modes have higher $\omega_{I}$ than $\omega_{R}$. This indicates that these modes are strongly damped.
In the same way the presence of $\omega_{I}>0$ would indicate instabilities of the scalar perturbation. However, if we choose $r_{0}^{+}$ in (\ref{bcrtcs13limit}) these modes are absent.

Using the boundary conditions (\ref{bcstcs1}), (\ref{bcstcs2}), (\ref{bcstcs3}) and (\ref{ctc}) in the large $m$ limit, equation (\ref{bcrtcs11}) is obtained with
\begin{equation}
\label{bcrtcs12largem}
R^{in}(r) = \tilde{C}_{1}\ P_{m}^{n}(\cos(\sqrt{\lambda}r))\quad,
\end{equation}
\begin{equation}
R^{out}(r) = \left\{
\begin{array}{lcl}
C_{1} H^{1}_{\nu}(p(r+r_{0})) & \quad \textrm{when} \quad & \omega_{R}>0 \\
\\
C_{2} H^{2}_{\nu}(p(r+r_{0})) & \quad \textrm{when} \quad & \omega_{R}<0
\end{array}
\right.
\quad ,
\end{equation}
where $n(n+1) = \left(\omega^2-k^2-\beta^2\right)/\lambda$. In Table \ref{tab4} some quasinormal modes for the rotating cosmic cylinder ``ballpoint pen" model in large $m$ limit are listed. The sign of the imaginary part $\omega_{I}$ depends again on the values of $r_{0}^{\pm}$. The presence of $\omega_{I}>0$ indicates instabilities of the scalar perturbation. On the other hand, if we choose $r_{0}^{+}$ in (\ref{bcrtcs13limit}) these modes are absent and we can declare the stability of the space-time under scalar perturbations. The precision of the approximation is higher in the first overtones. In this limit the field presents purely real modes of oscillation but they arise only when instabilities are also observed.

\subsubsection{Static and rotating cosmic string}

In the case of static and rotating string, the application of the boundary conditions (\ref{bcstcs2}) and (\ref{bcstcs3}) lead to Hankel functions. However when regularity condition at the origin is imposed, both Hankel functions diverge. Hence these space-times do not display quasinormal modes of oscillation.

%%%%%%%%%%%%%%%%%%%%%%%%%%%%%%%%%%%%%%%%%%%%%%%%%%%%%%%%%%%%%%%%%%%%%%%%%%%%%%%%
\section{Final Remarks}
%%%%%%%%%%%%%%%%%%%%%%%%%%%%%%%%%%%%%%%%%%%%%%%%%%%%%%%%%%%%%%%%%%%%%%%%%%%%%%%
\label{final}

The results in this paper concern the stability of cosmic cylinder space-times and its relation with closed time-like curves. The data obtained in previous sections show that, when the parameters are chosen in a such way that the scalar quasinormal modes have always $\omega_{I}<0$ then an inspection of the causal behavior reveals the absence of CTCs. On the other hand, if the parameters are chosen to guarantee the presence of CTCs in the space-time then the scalar quasinormal modes will present unstable modes $\omega_{I}>0$.

For the static cases and for the rotating cosmic strings and cylinders it is observed that the solutions can be divided in two groups: the ``group S'', where $r_{0}^{+}$ is chosen in Eqs.  (\ref{rctcballpoint}),(\ref{bcrtcs13}), (\ref{bcrtcs13limit}); and the ``group I'', where $r_{0}^{-}$ is chosen. Thus the constant $r_{0}^{\pm}$ controls the appearance of CTCs in these cosmic cylinder space-times. The results can be stated schematically as
\begin{center}
\begin{tabular}{|c|c|}
\hline
                     &                          \\
Scalar stability     &      Existence of CTCs   \\
$\omega_{I}<0$       &      $g_{\phi\phi}<0$     \\
                    &                           \\
$\Downarrow$        &         $\Downarrow$      \\
                    &                           \\
group S:            &        group I:           \\
\hspace{0.5cm}
 $r_{0}^{+}$ in Eqs. (\ref{rctcballpoint}) , (\ref{bcrtcs13}) ,
(\ref{bcrtcs13limit})
\hspace{0.5cm}
                    &
\hspace{0.5cm}
$r_{0}^{-}$ in Eqs. (\ref{rctcballpoint}) , (\ref{bcrtcs13}) ,
(\ref{bcrtcs13limit})  \hspace{0.5cm}
                                                \\
                    &                           \\
$\Downarrow$        &         $\Downarrow$      \\
                    &                           \\
Absence of CTCs     &    Scalar instability     \\
  $g_{\phi\phi}>0$   &        $\omega_{I}>0$     \\
                    &                           \\
\hline
\end{tabular}
\end{center}

Summarizing our results we conclude that cosmic cylinder space-times that present closed time-like curves are unstable against scalar perturbations.
Since the scalar field propagation gives us strong signals of instabilities in the considered space-times when the CTCs are present, our results for the scalar perturbation lead us to conjecture that the presence of closed time-like curves is one source of \textit{space-time} instabilities. The demonstration of this conjecture could open a new window in the investigation of causality.

\acknowledgments

The authors would like to thank FAPESP (Brazil) and CNPq (Brazil) by the financial support.

\end{document}